\documentstyle[prb,aps]{revtex}
\begin{document}
\draft

\title{Coherent phenomena in mesoscopic systems
\footnote{Invited talk presented at NATO ARW {\it "Supermaterials"} at
Giens, 18-23 September 1999}}
\author{E. Zipper and M. Lisowski}
\address{\it Institute of Physics, University of Silesia,
ul. Uniwersytecka 4, 40-007 Katowice, Poland}
\date{\today}
\maketitle

\begin{abstract}
A mesoscopic system of cylindrical geometry made of a metal or a
semiconductor is shown to exhibit features of a quantum coherent state.
It is shown that magnetostatic interaction can play an important role
in mesoscopic systems leading to an ordered ground state. The temperature
$T^{*}$ below the system exhibits long-range order is determined. The
self-consistent mean field approximation of the magnetostatic interaction
is performed giving the effective Hamiltonian from which the self-sustaining
currents can be obtained.

The relation of quantum coherent state in mesoscopic cylinders to other
coherent systems like superconductors is discussed.
\end{abstract}

\section{Introduction}

In the light of recent technological advances in nanostructure fabrication
there has been a renewed interest in the properties of quasi-one-dimensional
(quasi-1D) and two-dimensional (2D) electron systems. The transport
properties of mesoscopic metallic or semiconducting samples have been shown
to exhibit features characteristic of the quantum coherence of the electronic
wave function along the whole sample.

In this paper we want to discuss some of the properties of a quantum
coherent state of a mesoscopic system of cylindrical geometry.
We show that if we reduce the dimensions of a cylinder
made of a normal metal or a semiconductor to mesoscopic dimensions,
a system exhibits coherent properties absent in macroscopic samples.
We also discuss the possibility of long-range order (LRO) in a set of
mesoscopic rings deposited along $z$ axis.

We study the mesoscopic cylinders made of a normal metal or semiconductor
with quasi-1D and 2D conduction and we assume that electrons interact
via the magnetostatic (current-current) interaction.\\
The circumference, the height and the thickness of a cylinder are denoted
by $L_x =2\pi R$, $L_z$ and $d$ respectively. We assume that we have a thin
cylinder $d \ll L_x,L_z$ and that $L_x$ is of mesoscopic size.

Systems with quasi-1D conduction can be BCC crystals, low dimensional organic
conductors, and cylinders made of concentric mesoscopic rings stacked along
$z$ axis by e.g. lithographic method. Cylinders made of materials with
layered structure and multiwall carbon nanotubes are the examples of systems
with 2D conduction.

\section{Magnetostatic interaction in mesoscopic cylinders}

The properties of quasi-1D mesoscopic rings and 2D cylinders in the presence
of a static magnetic field are well known \cite{Cheung,IBM}. They exhibit in
the presence of a static magnetic flux $\phi_e$ persistent diamagnetic or
paramagnetic currents depending on $k_F$ and lattice constant $a$.
Moreover the magnetostatic interaction can lead to self-sustaining persistent
currents \cite{StebSzopZip,StebLisZip} even if we switch the external flux
off. Such currents are a hallmark of phase coherence.

In one of our recent papers \cite{LisZipSteb} we derived a microscopic
formula for the current-current interaction for a set of $M_z$ mesoscopic
rings stacked along $z$ axis. Then performing a self-consistent mean field
approximation (SMFA) we arrived at the effective Hamiltonian $H^{MF}$ from
which the self-sustaining currents can be obtained.
\begin{equation}
\label{m28}H^{MF}=\frac 1{2m_e}\sum_{m/1}^{M_z}\sum_{n/1}^{N_1}\left( p_{nm}
-eA_I\right) ^2+\frac{\phi _I^2}{2{\cal L}},
\end{equation}
where for long cylinders ($L_z \gg R$) ${\cal L}=\mu_0 \pi R^2 /L_z$;
$p_{nm}=p_{nm}^0 - eA_e$, $p_{nm}^0$ is a momentum of a $n$-th electron
in a $m$-th channel, $A_e = \phi_e /2\pi R$, $\phi_e$ is the external
magnetic flux; $\phi_I=2\pi RA_I$, $A_I$ is a vector potential coming from
all currents in the system; $N_1$ is the number of conducting electrons in
a single channel.

$H^{MF}$ was the basis of our previous investigations of spontaneous self-%
sustaining currents \cite{LisZipSteb,StebSzopZip}.\\
In this paper we want to check whether $H^{MF}$ given by Eq. (\ref{m28}) is
also valid for cylinders with 2D conduction.\\
The general formula for the magnetostatic interaction is of the form:
\begin{equation}
\label{m9}H_{mgt}=-\frac{\mu _0}{8\pi }\int d^3 {\bf r}d^3 {\bf r}%
^{\prime }\frac{{\bf J}({\bf r}) \cdot {\bf J}({\bf r}%
^{\prime })}{\left| {\bf r}-{\bf r}^{\prime }\right| },
\end{equation}
where ${\bf J}({\bf r})=e {\bf p}({\bf r})/m_e$, ${\bf J}({\bf r})$ is the
current density, ${\bf p}({\bf r})$ is the momentum of an electron.

Let us apply to our system a small magnetic field parallel to the $z$ axis.
The resulting currents will run in the $x$ direction in $M_z$ channels.\\
We obtain
\begin{equation}
\label{m11}H_{mgt}=-\frac 12\sum_{m/1}^{M_z}\sum_{m^{\prime }/1}^{M_z}{\cal L%
}_{mm^{\prime }}I_mI_{m^{\prime }}, 
\end{equation}
where $I_m$ is the current in the $m$-th channel,
\begin{equation}
\label{m12}{\cal L}_{mm^{\prime }}=\frac{\mu _0}{4\pi }\oint_{C_m}%
\oint_{C_{m^{\prime }}}\ \frac{d {\bf \xi}_m d {\bf \xi}_{m^{\prime }}}
{\left| {\bf \xi}_m- {\bf \xi}_{m^{\prime}}\right| },\qquad {\cal L}_
{mm^{\prime }}={\cal L}_{m^{\prime} m}.
\end{equation}
The interaction constant ${\cal L}_{mm^{\prime }}$ depends on the sample
geometry; here it has to be calculated for the cylinder with channels
at distance $z_{mm^{\prime }}=z_m-z_{m^{\prime }}$.

\ 

The $z$ dependence of the coupling constant ${\cal L}$ is presented
in Fig. 1. For small $z$ it falls down slowly proportionally to $\mu
_0R\left( \ln 8R/z-2\right) $, for large $z$ it falls down faster
proportionally to $1/z^3$. The interaction constant depends only on $R$
and on the relative distance of the channels. We see that the interaction
(\ref{m11}) is a long-range interaction. This indicates in particular that
thermodynamic fluctuations of the current will be strongly supressed
\cite{Fisher}.

If we express the currents $I_m$ via the momenta $p_m$, $H_{mgt}$ can be
rewritten in the form:
\begin{equation}
\label{m17}H_{mgt}=-\frac{e^2}{2m_e^2}\sum_{m/1}^{M_z}\sum_{m^{\prime
}/1}^{M_z}g_{mm^{\prime }}p_mp_{m^{\prime }}, 
\end{equation}
$$
g_{mm^{\prime }}=\frac 1{4\pi ^2 R^2}{\cal L}_{mm^{\prime }},
$$
where
\begin{equation}
\label{m16}I_m=\frac e{2\pi Rm_e}p_m, 
\end{equation}
$$
p_m=\sum_{n/1}^{N_1}p_{nm}.
$$

Let us perform a self-consistent MFA of the interaction (\ref{m17}), such
approximation is known to be good for a long-range interaction.
\begin{equation}
\label{m22}H_{mgt}^{MF}=-\frac e{2m_e}\sum_m\left[ 2p_m\left\langle
A_I(z_m)\right\rangle -\left\langle p_m\right\rangle \left\langle
A_I(z_m)\right\rangle \right] ,
\end{equation}
where the first term in Eq. (\ref{m22}) has been obtained by use of the
symmetry relation (\ref{m12}), $\left\langle A_I(z_m)\right\rangle
=(e/2m_e)\sum_{m^{\prime }}g_{mm^{\prime }}\left\langle p_{m^{\prime
}}\right\rangle \equiv A_I$.\\
For the cylinder geometry the formula for $A_I$ takes the form:
\begin{equation}
\label{m23}A_I=\mu _0R\frac{\left\langle \sum_{m/1}^{M_z} I_m\right\rangle }
{2L_z}.
\end{equation}
Calculating the current $\left\langle \sum_{m/1}^{M_z} I_m\right\rangle $
with a total vector potential $A=A_e+A_I$ we obtain after some algebra
\begin{equation}
\label{m27}
eA_I=\frac \eta {1+\eta N_1 M_z} \sum_{m/1}^{M_z} \left\langle
p_m \right\rangle ,
\end{equation}
where
$$
\eta =\frac{\mu _0e^2}{4\pi L_z m_e}.
$$
Inserting Eq. (\ref{m27}) into (\ref{m22}) and adding to $H_{mgt}^{MF}$ the
kinetic energy term we obtain the Hamiltonian $H^{MF}$ given by Eq.
(\ref{m28}).

Thus we proved that the SMFA of the microscopic Hamiltonian given by Eq.
(\ref{m28}) is also valid for systems with 2D conduction. However $A_I$
and hence $\phi_I$ has to be calculated in a different way than in the
systems with quasi-1D conduction \cite{LisZipSteb}.\\
The basic difference between systems with quasi-1D and 2D conduction is the
formula for the total current $I$.
Systems with quasi-1D conduction have flat Fermi surfaces (FS) perpendicular
to $k_x$ direction. There exists then a largest correlation among the
currents from different channels (rings) and the total current $I$ is the
biggest. In the 2D case the current $I$ depends strongly on the shape of
the FS.

The total current $I$ in the system can be written as a Fourier series
\cite{LisZipSteb}:
\begin{equation}
\label{AA}
I(\phi)=M_r \sum_{m/1}^{M} \sum_{g/1}^{\infty} \frac{4I_0(m)}{\pi}
\left( \frac{L_x}{2\gamma}+\frac{2\pi ^2 k_BT}{\Delta_0}\right) \frac{\exp
\left[ -g\left( \frac{L_x}{\gamma}+\frac{2\pi^2 k_BT}{\Delta_0}\right)
\right] }{1-\exp\left[ -g\left( \frac{L_x}{\gamma}+\frac{4\pi^2 k_BT}
{\Delta_0}\right) \right] }\cos \left[ g k_{F_x}(m) L_x \right]
\sin \left( 2\pi g \frac{\phi}{\phi_0} \right) ,
\end{equation}
where $\phi_0 = h/e$, $1/\gamma$ is a disorder parameter \cite{LisZip},
$\Delta_0 = \hbar ^2 N_1 /(2m_e R^2)$ is the quantum size energy gap at
the FS for electrons running in the $x$ direction, $I_0(m)=e\hbar k_{F_x}(m)
/m_eL_x$, $k_{F_x}(m) = k_F \left[ 1-\left( k_{F_z}(m)/k_F \right) ^u
\right] ^{1/u}$, $k_z(m)=m \pi / L_z$, $M$ is the number of channels in the
$k_z$ direction, $M_r$ is the number of channels on the thickness $d$.
\cite{LisZipSteb}
\\
$u$ is the parameter used to model different shapes of the FS. \cite{Crack}
Changing $u$ from 2 to $u \gg 2$ we can investigate the shape of the FS
from circular to rectengular with rounded corners, with different amount
of flat regions. Currents given by Eq. (\ref{AA}) are persistent at $k_B T
\ll \Delta_0$ because the energy gap prevents scattering.

The flux $\phi$ which drives the current is the sum of the external flux
$\phi_e$ and the flux $\phi_I$ from the current itself,
\begin{equation}
\label{BB}
\phi = \phi_e + \phi_I , \qquad \phi_I = {\cal L}I.
\end{equation}
The self-consistent solutions of Eqs. (\ref{AA}) and (\ref{BB}) at $\phi_e =
0$ give the values of self-sustaining flux. They are presented as circles
in Figs. 2, 3 for a quasi-1D system and in Fig. 4 for a 2D system.\\
In Fig. 2 the case of a system running paramagnetic persistent currents
is considered. The self-sustaining solutions correspond to spontaneous fluxes
and are obtained for temperature $T<T_c = 3.5$ K, $T_c$ is the transition
temperature to an ordered state.

The system discussed in Fig. 3 exhibits strong (Meissner-like) diamagnetic
reaction at small fluxes. The self-sustaining solutions correspond to trapped
flux (for $T<T_c = 0.6$ K). We see that the coherent state of a system
presented in Fig. 3 posses features characteristic of superconductors.\\
We notice that the spontaneous flux solution can be obtained at higher
temperatures than the flux trapped solution because the curves from Eqs.
(\ref{AA}) and (\ref{BB}) have to cross in the first and the second half
of the $\phi /\phi_0$ period, respectively.

The current-flux characteristics for 2D cylinders with different shapes of
the FS are presented in Fig. 4. We see that the self-sustaining current
solutions can be obtained only for $u \geq 5$, i.e. for systems with the FS
having substantial flat regions. Such FS are frequently met, e.g. in high
temperature superconductors \cite{King}.

\section{Effective long-range order}

Having the microscopic $H$ of the magnetostatic interaction [Eq. (\ref{m17})]
we can discuss the concept of effective long-range order (LRO) and phase
transitions in finite quasi-1D systems. Let us focus on a set of stacked
rings. As the currents $I_m$ can run only in the clockwise or anticlockwise
direction the Hamiltonian (\ref{m11}) has the form  of the Ising-like
Hamiltonian. It has been well established that 1D systems with short range
forces cannot have a phase transition at finite in the thermodynamic limit
\cite{Imry}. For a system described by an Ising Hamiltonian $H_I$ the proof
goes as follows
\begin{equation}
\label{O}
H_I=-\sum_{m/2}^{M_z}\sum_{n/1}^{m-1}J(m-n)S_mS_n
\end{equation}
In the ground state all "spins" are parallel $(S_mS_n=\pm 1)$. The change in
the free energy $\Delta F$ connected with reversing the direction of $L$
"spins" $\left( L\ll M_z \right) $ in $p$ $\left( 1\leq p\ll M_z \right) $
different places is
\begin{equation}
\label{A}
\Delta F=p\Delta E_L-T\Delta S=p\left( \Delta E_L-k_B T\ln M_z \right) ,
\end{equation}
where $\Delta E_L$ is the change in the internal energy.

For short-range (n.n) interactions $\Delta E_L=2J$ is independent of $M_z$
and $L$ and $\Delta F<0$ for $M_z \rightarrow \infty $. It means that the
configuration with domain structure has always lower energy in the
thermodynamic limit and there is no LRO. However for finite $M_z$ long-range
order is effectively obtained if $\Delta F>0$. We can define a correlation
range
\begin{equation}
\xi_T=e^{\Delta E_L/k_BT},
\end{equation}
and we see from Eq. (\ref{A}) that the system has an ordered ground state
if $\xi_T > M_z$, i.e. if the correlations extend over all length.

The microscopic Hamiltonian of the magnetostatic interaction given by
Eq. (\ref{m11}) is of the form of an Ising Hamiltonian but the situation is
much more interesting here because the interaction is long-range.\\
The energy change when reversing a direction of $L$ currents is given by
\begin{equation}
\label{C}
\Delta E_L (M_z) = \frac{1}{2}\left( \frac{e\hbar N_1}{2\pi m_eR^2}\right)
 ^2 \sum_{n/1}^{L} \sum_{m-m'/L+1}^{M_z-n} {\cal L}(m-m').
\end{equation}
The inspection of $\Delta E_L (M_z)$ for different values of
$b=z_{m+1}-z_{m}$ shows that it first increases with increasing both
$L$ and $M_z$ and then saturates. From $\Delta F=0$ we can calculate
the temperature $T^*$ at which the crossover from an ordered to disordered
state occurs
\begin{equation}
\label{D}
T_L^* =\frac{\Delta E_L(M_z)}{k_B\ln M_z}.
\end{equation}
We can investigate now the possibility of LRO in a set of $M_z$
$(M_z \gg 1)$ stacked rings. We discuss different possibilities.

At first we can consider the instability against the formation of long
domains ($L\gg 1$) of rings with currents running in the opposite directions.
Taking e.g. $L=1000$ and $b=60$ \AA, $R=5000$ \AA, $M_z=10^4$ we get from
Eqs. (\ref{C}) and (\ref{D}) $T_{1000}^{*}\sim 1.5$ K.\\
Thus at $T<1.5$ K the instability implied by Eq. (\ref{D}) excludes the
existance of disorder in the form of long alternating domains with opposite
currents. However short-range disorder with small number of rings with
reversed currents is not excluded.

In order to get such long-range stable domains one should need some
additional interaction \cite{Imry} which should stabilize long domains
against short-range fluctuations. Then with such additional mechanism
included, one could say that at $T<T^*$ the system exhibits long range order.

Let us discuss now the instability against short-range disorder.
The smallest energy change is obtained when reversing a direction of a single
current in several places. Taking $L=1$ and other parameters the same as in
previous example we obtain from Eqs. (\ref{C}) and (\ref{D})
$$
T_1^{*}\sim 2.2\mbox{ x }10^{-2} \mbox{ K}.
$$
However we can also assume that stacked rings have finite but small
thickness $d$. Then the number of channels in each ring increases what
results in increasing $\Delta E_1$ and hence $T_1^{*}$.\\
Taking e.g. $d=10$ \AA, $b=60$ \AA, $R=5000$ \AA, $L=1$, $M_z=10^4$ we obtain
from Eq. (\ref{D}) $T_1^{*}\sim 0.45$ K,\\
and for $d=20$ \AA, $b=110$ \AA, other parameters as above, we obtain
$T_1^{*}=0.98$ K.\\
Finite values of transition temperatures are related to finite size of the
considered samples. Indeed if we put $M_z \rightarrow \infty$ then due to
the fact that $\Delta E_L (M_z)$ saturates for large $M_z$, $T_L^{*}
\rightarrow 0$.

It has been proved many years ago \cite{Ellis,Frohlich} that for the
interaction $J(m-n) = J / \left| m-n \right| ^\alpha$, $\alpha >2$ in
Eq. (\ref{O}) the model exhibits no phase transition at finite temperature.
In our case ${\cal L}(m-n) \sim 1 / \left| m-n \right| ^3$ for large
$\left| m-n \right| $ and we do not expect LRO in the thermodynamic limit.
So our system is an example when large but finite system can show interesting
effects which will be wiped away in the thermodynamic limit.\\
We want to stress that such conclusions are valid only if the systems can be
treated with good approximation as 1D systems. The systems considered by us
in the previous section were 2D systems and they should be described by a 2D
Ising model which has a phase transition to an ordered state in the
thermodynamic limit. Such consideration will be a subject of our further
work.

\section{Conclusions}

In the presented paper we discussed some aspects of quantum coherence in
mesoscopic systems. One of the basic thermodynamical properties of a
macroscopic metal or semiconductor is a very weak reaction to a static
magnetic field (weak Landau diamagnetism).

Whether or not a sample of a given size can be considered as a macroscopic
one, depends on temperature. It turns out that if $k_B T< \Delta_0$
the system exhibits mesoscopic orbital magnetism \cite{Shapiro} which
manifests itself in strong paramagnetic or diamagnetic persistent currents
in the presence of a static magnetic field. What is more, if the magnetic
response is sufficiently large one should account also for the field produced
by the orbital currents and solve the entire problem self-consistently.\\
To treat this problem we considered the long-range magnetostatic interaction.
Performing the SMFA we arrived at the effective Hamiltonian, valid both in
systems with quasi-1D and 2D conduction which leads to self-consistent
equations for the current.

The self-sustaining solutions (at $\phi_e = 0$) for samples exhibiting
paramagnetic behavior at low $\phi$ are spontaneous currents manifesting
a break of time reversal symmetry.\\
The self-sustaining solutions (at $\phi_e = 0$) for samples exhibiting
diamagnetic behavior at low $\phi$ are equivalent to flux trapped in the
systems. Such behavior was previously attributed solely to superconductors,
we show here that it can be obtained in a coherent state in mesoscopic
cylinders.\\
The self-sustaining currents present in mesoscopic systems with low-%
dimensional conduction, are a hallmark of a quantum coherence.
The amount of coherent electrons running persistent currents is very
sensitive to the shape of the FS. The most favorable situation is for the
system with flat FS where the correlation of currents from different channels
is the strongest.

The microscopic $H$ of the magnetostatic interaction for a cylinder made of
a set of stacked rings has the form of an Ising Hamiltonian with the
long-range interaction. Using it we have discussed the effective long-range
order for large but finite systems. We calculated the temperature $T^{*}$
below which the system is completly ordered in a sense that the correlation
range extends over its length.
Most of the theorems discussing LRO are valid only in the thermodynamic
limit. They therefore miss some interesting properties of the system in
question.

The properties of such quasi-1D systems are interesting by itself
but can also be used to study phase transitions in materials which can be
viewed as arrays of quasi-1D systems. This will be studied in a forthcoming
paper.

\newpage

{\bf Figure captions}

\begin{figure}
\caption{The interaction constant ${\cal L}$ as a function of distance $z$
between channels.}
\end{figure}

\begin{figure}
\caption{Paramagnetic persistent currents $I/I_0$ as a function of a magnetic
flux $\phi / \phi_0$ at $\phi_e=0$ and different $T$. Spontaneous flux
solutions are denoted by circles.}
\end{figure}

\begin{figure}
\caption{Diamagnetic persistent currents $I/I_0$ as a function of a magnetic
flux $\phi / \phi_0$ at $\phi_e=0$ and different $T$. Trapped flux solutions
are denoted by circles.}
\end{figure}

\begin{figure}
\caption{Persistent currents $I/I_0$ as a function of magnetic flux $\phi /
\phi_0$ at $\phi_e=0$ in 2D mesoscopic cylinders with different shapes of
the Fermi surfaces. Self-sustaining currents are denoted by circles.
$N$ is a number of conducting electrons in a single cylinder.}
\end{figure}

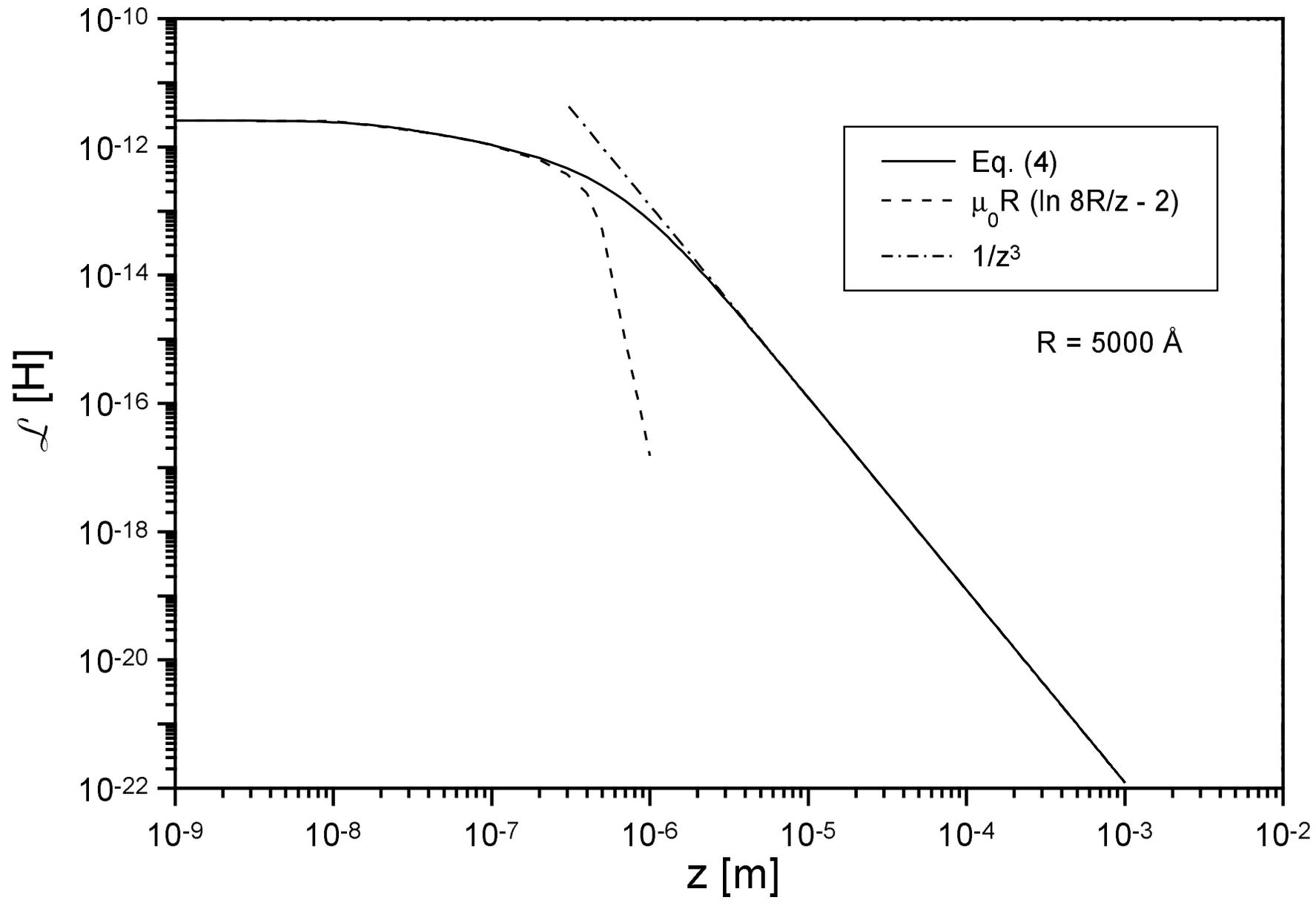
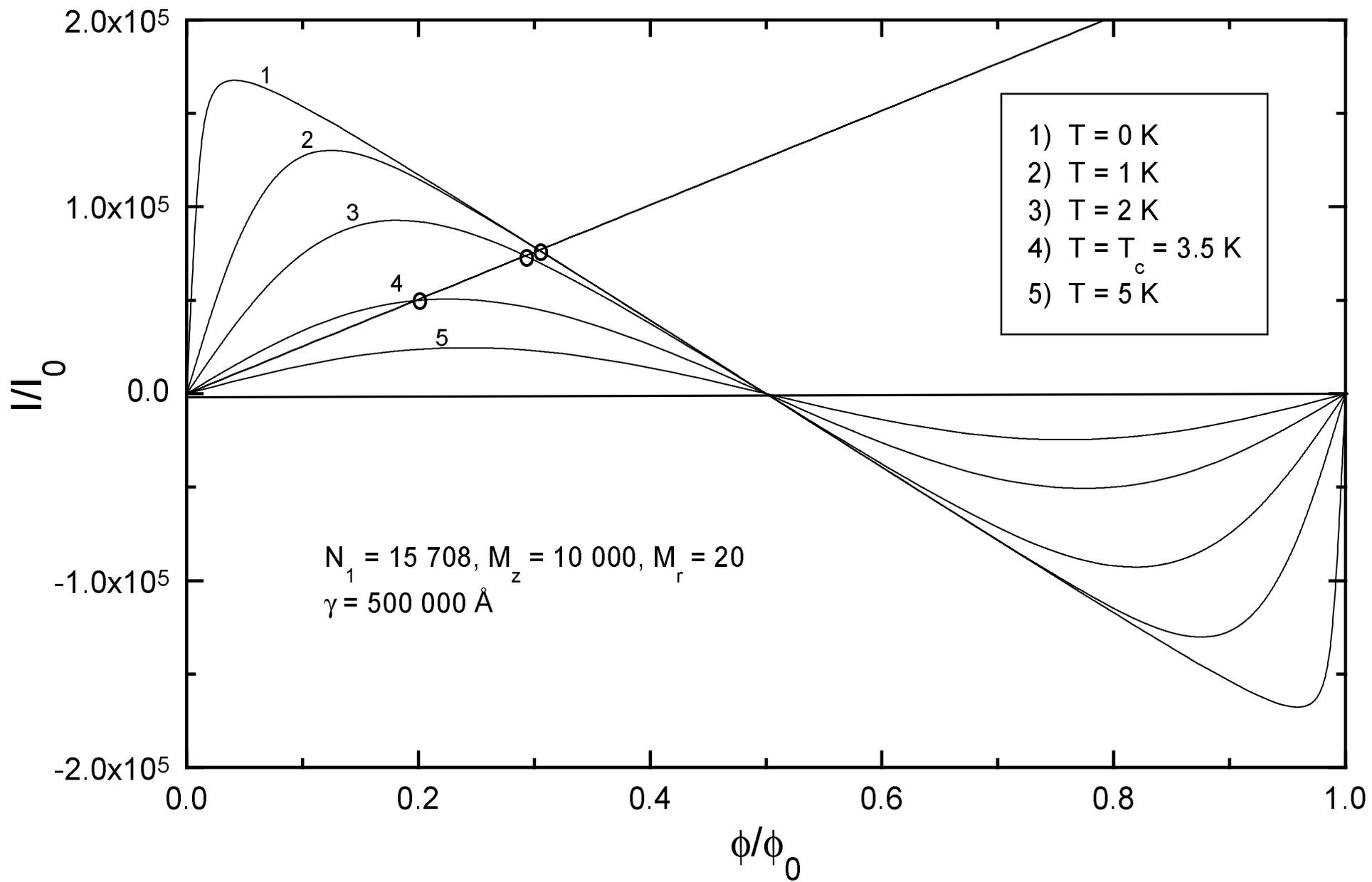
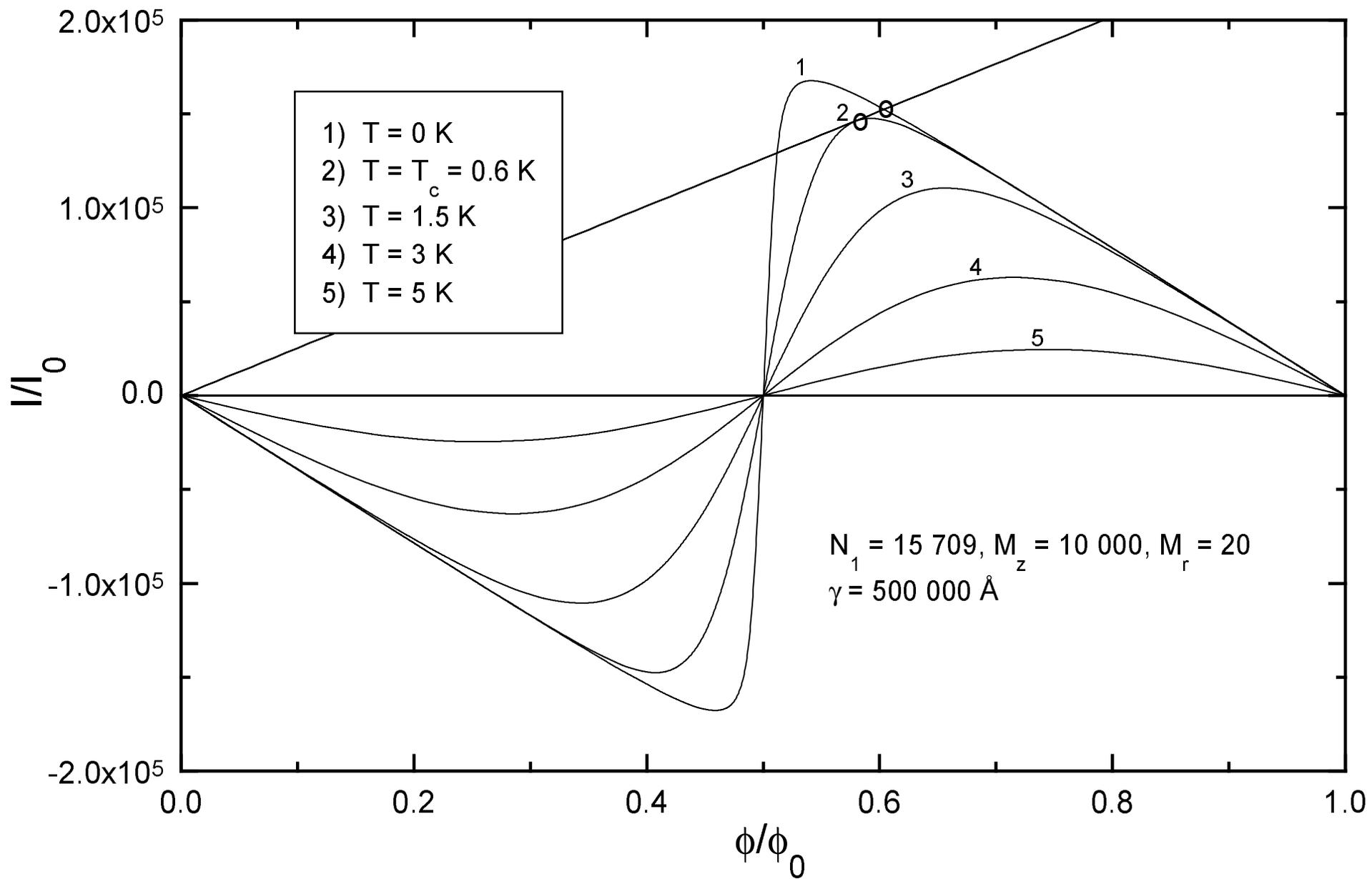
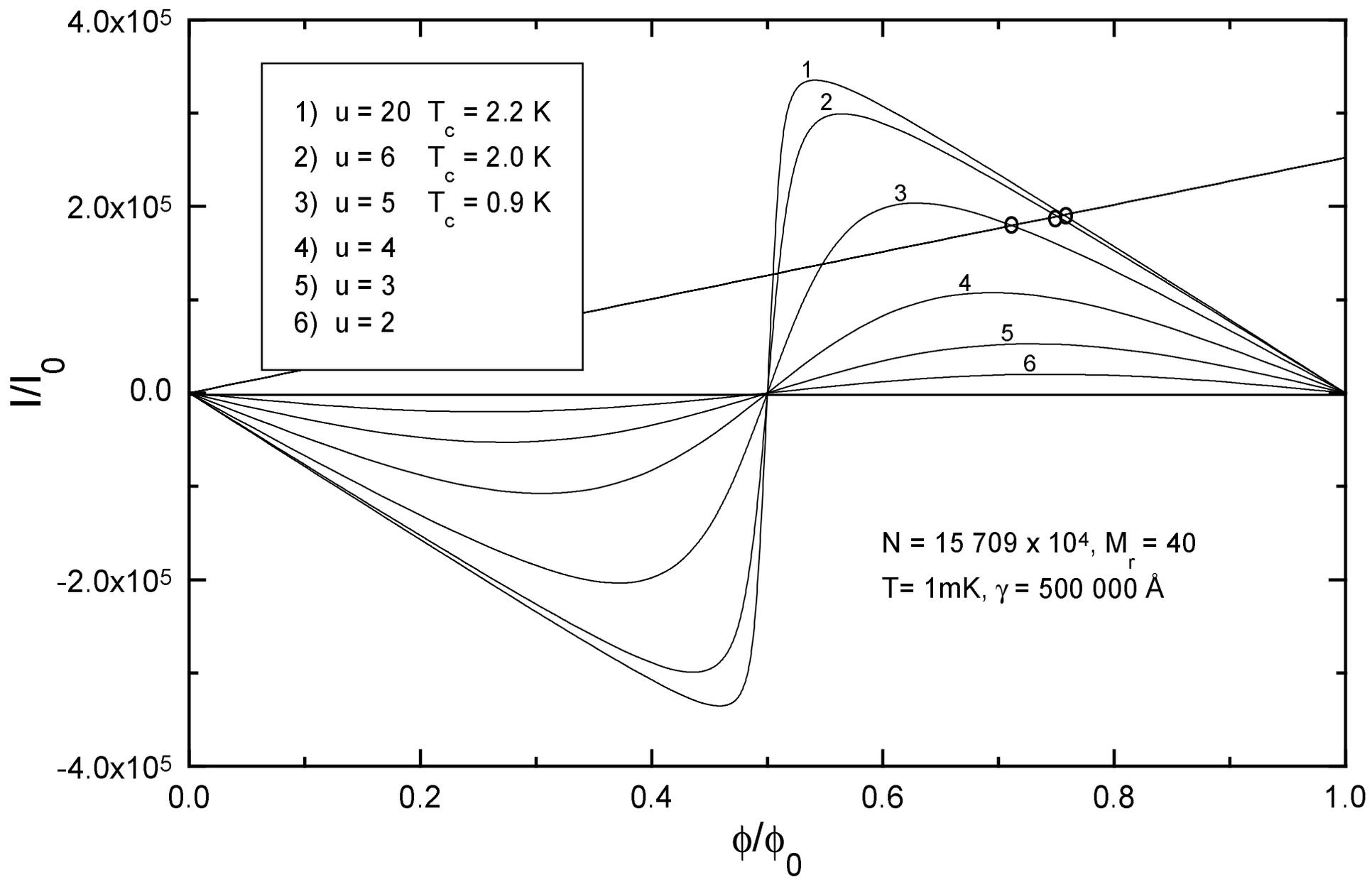
\end{document}